\documentclass[twocolumn,pra]{revtex4}
\usepackage{amsmath}
\usepackage{amssymb}
\usepackage{graphicx,psfrag,color}
\usepackage{dcolumn}
\usepackage{bm}
\usepackage{mathbbol}
\usepackage{amsthm}
\usepackage{multirow}

\makeatletter
\@addtoreset{equation}{section}
\makeatother

\begin{document}

\title{Hybrid direct state tomography by weak value}

\author{Xuanmin Zhu{$^{1,2}$}}\email{zhuxuanmin2006@163.com}
\author{Qun Wei$^3$}
\author{Lixia Liu$^4$}
\author{Zijiang Luo$^1$}

 \affiliation{$^{1}$ School of Information, Guizhou University of Finance and Economics, Guiyang, Guizhou 550025, China\\
 $^{2}$ International Joint Research Center for Data Science and High-Performance Computing, School of Information, Guizhou University of Finance and Economics, Guiyang, Guizhou 550025, China\\
 $^{3}$ School of Physics and Optoelectronic Engineering, Xidian University, Xi'an 710071, China \\
 $^{4}$ School of Mathematics and Statistics, Xidian University, Xi'an 710071, China}

\begin{abstract} \normalsize
Compared with the conventional quantum state tomography (QST), the direct state tomography (DST) using weak value is easily manipulated in experiments. However, the efficiency of the DST is lower than that of the conventional QST, especially for high-dimensional systems. For a pure state, the DST is revised to improve the efficiency.  In the revised DST, the real or imaginary parts of the weak values can be obtained by measuring only one system observable. We constructed a hybrid DST by combining the original DST and the revised DST. By using the appropriate measurement strength, the efficiency of the hybrid DST is significantly larger than that of the conventional QST, especially for high-dimensional systems. The state reconstruction strategy investigated in this paper may be useful in actual experiments.
\end{abstract}

\maketitle

\section{Introduction}\label{sec1}
The unknown quantum state can not be perfectly cloned and determined by measuring only one quantum system~\cite{niel}. In quantum information theory, quantum state tomography (QST) which is used to reconstruct the unknown state by measuring the identical quantum systems is an important topic~\cite{qst001,qst002,qst1,qst2,qst3,qst4,qst5,qst6,qst7,qst8, qst9, qst10, qst11, qst12, qst13, qst14,qst15}. In the conventional QST, it requires the measurements of a complete set of noncommutative observables, which are difficult to be realized perfectly in actual experiments, especially for high-dimensional systems.

A direct state tomography (DST) is proposed by using the concept of weak value in weak measurements~\cite{aav, wk1, wk2, wk3, wk4, dsm1, dsm2,dsm3,dsm4,dsm5}. In DST, the elements of the unknown density matrix can be directly reconstructed by the results of the weak values, which can be achieved just by measuring two observables of the measuring devices. DST is much more easily realized compared to the conventional QST, which has attracted much attention, especially, in the reconstructing the high-dimensional unknown quantum states. By using the method of DST, one-million-dimensional and 19200-dimensional state reconstructions have been realized~\cite{dsm6,dsm7}.

In DST, for a $d$-dimensional quantum system, only $1/d$ elements of the real (or imaginary) part of the density matrix can be obtained in each measurement. To accomplis $N$ times measurements, the number of the unknown quantum systems we needed is $2dN$. This is the reason why DST is less efficient than the conventional QST~\cite{mac,zhan,gro,vall,zou,zhu1,zhu2}. Could we revise the DST to make its efficiency higher than that of the conventional QST?

In this paper, we proposed a revised DST in which we could obtain the real (imaginary) part of the weak value by measuring only one observable for a pure unknown state. In the revised DST, we can accomplish $N$ times measurements by using only $2N$ identical quantum systems, which is independent on the dimension of the system. The efficiency of the revised DST is significantly improved for some states, but the revised DST is useless for other states. To overcome the useless, we construct a hybrid DST by combining the original DST and revised DST. By the Monte Carlo simulations, we will show that the efficiency of the hybrid DST is significantly higher than that of the conventional QST.

This paper is organized as follows. We reviewed the original DST in Sec.~\ref{sec2}, and proposed a revised DST in Sec.~\ref{sec3}. In Sec.~\ref{sec4}, we constructed the hybrid DST, and compared the efficiency of the hybrid DST with that of the convention QST in Sec.~\ref{sec5}. A short conclusion is given in Sec.~\ref{sec6}.

\section{Direct state tomography}\label{sec2}
In this section, we simply review the original DST which is based on the weak measurement theory. It must be pointed out that the original DST in this article is the one named MDST (modified direct state tomography) in Ref.~\cite{zhu1}. The procedure of the original DST is as follows. First, we perform a measurement on the unknown system $\rho_s$ with some fixed measurement strength. The observable being measured of the system is one of a set of basis projectors $\{A_n=|a_n\rangle\langle a_n|\}$. Second, we implement a strong projective measurement on the system along another basis $\{|\psi_j\rangle\}$ and record the shifts of the measuring device and the obtained final system states, a procedure usually called postselection. Third, from the readings of the measuring device we can obtain the \emph{weak value} $W_{nj}$ which is defined as~\cite{dsm3}
\begin{equation}\label{e1}
W_{nj}=\frac{\langle \psi_j|a_n\rangle\langle a_n |\rho_s|\psi_j\rangle}{P_j},
\end{equation}
where $P_j$ is the probability of obtaining the final state $|\psi_j\rangle$. From Eq. (\ref{e1}), we have
\begin{equation}\label{e2}
\langle a_n|\rho_s|a_m\rangle=\sum_{j}P_j\frac{\langle \psi_j|a_m\rangle}{\langle \psi_j|a_n\rangle} W_{nj}.
\end{equation}
We can see that the matrix elements $\{\rho_{nm}\}$ of the unknown state $\rho_s$ can be reconstructed by using the values of the weak values $\{W_{nj}\}$. Thus, the key point for the DST is the measurement of the weak values.

In DST, independent of the dimension of the measured quantum system, the measuring device can be a pure two-dimensional system. Without loss of generality, the initial state of the measuring device could be denoted as
\begin{equation}\label{e3}
\rho_d = |0\rangle_d\langle 0|,
\end{equation}
where $|0\rangle_d$ is the eigenstate of $\sigma_z$ with the eigenvalue $1$. In DST, if the measured observable the system is $A_n=|a_n\rangle\langle a_n|$, the impulse interaction Hamiltonian between the system and measuring device can be described as
\begin{equation}\label{e4}
H_n=g\delta(t-t_0)A_{n}\otimes \sigma_x,
\end{equation}
where $g$ is the coupling strength. The combine state of the system and measuring device evolves through the unitary transformation $U_n=e^{-ig |a_n\rangle\langle a_n|\otimes \sigma_x}$, with $\hbar=1$ in this paper. Conditioned on obtaining the final state $|\psi_j\rangle$, the final measuring device state is
\begin{equation}\label{e5}
\rho_d^{nj}=\frac{\langle \psi_j|U_n\rho_s\otimes\rho_dU_n^{\dagger}|\psi_j\rangle}{P_j},
\end{equation}
where $\rho_d=|\Phi\rangle\langle\Phi|$ is the initial pointer state, and $P_j$ is the probability of obtaining $|\psi_j\rangle$. When the coupling strength is weak $g\to 0$, ignoring the terms $O(g^2)$, we have a approximate state
\begin{equation}\label{e6}
\rho_d^{nj}\approx\widetilde{\rho}_d^{nj}=\frac{\langle \psi_j|\rho_s|\psi_j\rangle}{P_j}\rho_d-ig\left(W_{nj} |1\rangle_d\langle 0|-W_{nj}^* |0\rangle_d\langle1|\right),
\end{equation}
where $\widetilde{\rho}_d^{nj}$ is used to distinguish the exact state $\rho_d^{nj}$. The weak value can be obtained by the expectation values of $\sigma_y$ and $\sigma_x$ under the approximate pointer state $\widetilde{\rho}_d^{nj}$, which can be described by the formula
\begin{equation}\label{e7}
W_{nj}=\frac{1}{2g}\left[-\mathbf{Tr}(\widetilde{\rho}_d^{nj} \sigma_y) +i\mathbf{Tr}(\widetilde{\rho}_d^{nj}\sigma_x)\right].
\end{equation}
The above equation is approximately valid when the coupling strength is weak. Thus, the efficiency of the DST based Eq. (\ref{e7}) is low, and there is a system error in the DST as shown in Refs.~\cite{mac,zhu1}.

In order to overcome these two drawbacks, we can use the coupling-deformed pointer observables to measure the weak values without approximation as proposed in Refs.~\cite{zhan,vall,gro,zou,zhu1,zhu2}. The corresponding operators which are dependent on the coupling strength $g$ of $\sigma_y$ and $\sigma_x$ are
\begin{equation}\label{e8}
\begin{split}
\sigma_y'(g)&=\frac{g}{\sin g}\left[\sigma_y-\tan \left(\frac{g}{2}\right)\left( I-\sigma_z \right)\right],\\
\sigma_x'(g)&=\frac{g}{\sin g}\sigma_x.
\end{split}
\end{equation}
Instead of Eq. (\ref{e7}), we can obtain the exact weak values by measuring the above two operators without any approximation from the formula
\begin{equation}\label{e08}
W_{nj}=\frac{1}{2g}\left[-\mathbf{Tr}(\rho_d^{nj} \sigma_y') +i\mathbf{Tr}(\rho_d^{nj}\sigma_x')\right].
\end{equation}
Substituting Eq. (\ref{e08}) into Eq. (\ref{e2}), we have
\begin{equation}\label{e9}
\langle a_n|\rho_s|a_m \rangle=\frac{1}{2g}\sum_{j}P_j\frac{\langle \psi_j|a_m\rangle}{\langle \psi_j|a_n\rangle} \left[-\mathbf{Tr}(\rho_d^{nj} \sigma_y')+i\mathbf{Tr}(\rho_d^{nj}\sigma_x')\right].
\end{equation}
Therefore, the elements of an unknown density matrix $\rho_s$ can be reconstructed directly by the expectation values of $\sigma_y'$ and $\sigma_x'$ under the exact state $\rho_d^{nj}$. In the strategy described by Eq. ({\ref{e9}}), there is no system error in the reconstructed state. For convenience, the bases $\{|a_n\rangle\}$ and $\{|\psi_j\rangle\}$ are always chosen as the mutually unbiased bases (MUBs)~\cite{mub}, which satisfy $\langle \psi_j|a_n\rangle=e^{2\pi jn i/d}/\sqrt{d}$.

Eq. (\ref{e9}) is valid for arbitrary coupling strength. We can improve the efficiency of the DST by choosing the optimal coupling strength. As shown in Ref.~\cite{zhu1,zhu2}, the efficiency of the DST is improved significantly. However, the improved efficiency of the DST is still lower than that of the conventional QST, especially for high-dimensional quantum systems~\cite{zhu1,zhu2}. In Eq. (\ref{e9}), it can be seen that we can obtain only one column of the real or imaginary part of the density matrix for once measurement. Thus, we need $2dN$ identical unknown systems to accomplish $N$ times measurements for $d$-dimensional system. This is the reason why the efficiency of the DST is low. Could we overcome this defect to improve the efficiency? In next section, we will propose a strategy in which $N$ times measurements could be accomplished by using $2N$ identical unknown systems for pure state.
\section{Revised direct state tomography}\label{sec3}

In this section, in order to improve the efficiency, we will revise the direct state measurement strategy for pure state. If the unknown state is a pure $\rho_s=|\phi_s\rangle\langle\phi_s|$, from Eq. (\ref{e1}), we have
\begin{equation}\label{e010}
\langle \psi_j|\phi_s\rangle=\left(\frac{P_jW_{nj}}{\langle\psi_j|a_n\rangle\langle a_n|\phi_s\rangle}\right)^*.
\end{equation}
The unknown state could be rewritten by
\begin{equation}\label{e011}
|\phi_s\rangle=\frac{1}{\langle\phi_s|a_n\rangle}\sum_j \left(\frac{P_jW_{nj}}{\langle\psi_j|a_n\rangle}\right)^*|\psi_j\rangle,
\end{equation}
where $P_jW_{nj}$ could be obtained by measuring only one operator $A=|a_n\rangle\langle a_n|$ as shown in Eq. (\ref{e08}). The states $\{|a_n\rangle\}$ and $\{|\psi_j\rangle\}$ are always chosen as mutually unbiased~\cite{mub}, which satisfy $|\langle \psi_j|a_n\rangle|=1/\sqrt{d}$ for each pair of $(n,j)$. The unknown common factor ${1}/{\langle\phi_s|a_n\rangle}$ could be removed by normalization. The unknown density matrix could be reconstructed by
\begin{equation}\label{e012}
|\phi_s\rangle\langle\phi_s| =\frac{1}{|\langle\phi_s|a_n\rangle|^2}\sum_{jk} \left(\frac{P_jW_{nj}}{\langle\psi_j|a_n\rangle}\right)^*\frac{P_kW_{nk}}{\langle\psi_k|a_n\rangle}|\psi_j\rangle\langle \psi_k|.
\end{equation}
From Eq. (\ref{e012}), we can see that the unknown pure state could be reconstructed by measuring only one operator $|a_n\rangle\langle a_n|$, which is independent of the dimension of the system. In order to distinguish the original DST, we use "revised DST" denotes this strategy based on Eq. ({\ref{e012}}). In the revised DST, we can accomplish $N$ times measurements using only $2N$ identical systems.

In order to gauge the efficiency of the tomography, we use the mean-square error (MSE) to measure the discrepancy between the true state $\rho$ and the reconstructed state $\rho_r$~\cite{qst11,qst12,qst13}, and the MSE is defined as
\begin{equation}\begin{split}\label{e013}
\mathcal{E}(\rho)&\equiv E (\Vert \hat{\rho}-\rho \Vert^2_{HS}),\\
&=\frac{1}{N_0}\left[\mathrm{tr}\left(E(\hat{\rho}^{\dagger}\hat{\rho})\right)-\mathrm{tr}(\rho^2)\right],
\end{split}\end{equation}
where $\hat{\rho}$ is the estimator and $N_0$ is the number of the copies. In DST, when the number of the copies is large, the MSE could be calculated approximately by
\begin{equation}\label{e014}
\mathcal{E}(\rho)\approx \mathrm{tr}(\rho_r^{\dagger}\rho_r)-\mathrm{tr}(\rho^2).
\end{equation}
The less MSE means the higher efficiency. We find that the MSE of the revised DST is dependent on the input state.

For two-dimensional systems, we perform the Monte Carlo simulations to show the dependence. In the simulations, the postselected states is $\{1/\sqrt{2}(|0\rangle +|1\rangle), 1/\sqrt{2}(|0\rangle-|1\rangle)\}$. The unknown state is denoted by $|\phi_s\rangle= \cos(\theta/2)|0\rangle+\sin(\theta/2)|1\rangle$. In order to eliminate the influence
of the statistical fluctuations, we average the values of MSEs over $10^4$ repeated simulations.

The simulation results have been presented in Fig. 1. The MSE of the revised DST is dependent on the unknown state as shown in Fig. 1. We can also see that the efficiency is high when $\langle \phi_s|A|\phi_s\rangle\to 1/d$; while the tomography strategy is useless when $\langle \phi_s|A|\phi_s\rangle\to 0$. The useless and low efficiency for some states makes the revised DST impractical. In the next section, we will propose a new strategy to overcome the defect of the revised DST.

\begin{figure}[t]
\centering \includegraphics[scale=0.5]{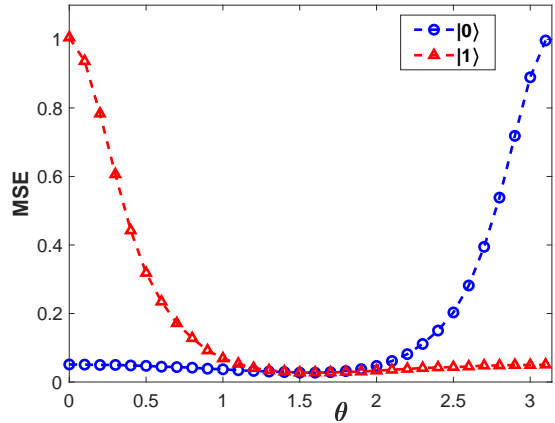} \caption{(Color online) The values of MSEs are simulated for different unknown pure states, which are averaged over $10^4$ repeated reconstructions of the revised DST. As the input unknown states is expressed as $|\phi_s\rangle= \cos(\theta/2)|0\rangle+\sin(\theta/2)|1\rangle$, the value of MSE is large when $\langle \phi_s |A|\phi_s\rangle \to 0$, the MSEs attain the small values when $\langle \phi_s |A|\phi_s\rangle \to 1/d$. The number of the unknown quantum systems is 100, and the measurement strength is $g=1.2$.}
\label{fig:1}
\end{figure}

\section{Hybrid direct state tomography}\label{sec4}
In this section, to eliminate the useless and the low efficiency of the revised DST for some unknown states, a hybrid DST is constructed by combining the original DST and the revised DST. In the hybrid DST, the measurements are divided two steps. The total $N$ unknown systems are divided into two parts, the number of one part is $N_1$, and the number of the other part is $N_2$. In the first step, we reconstruct a estimated pure state $|\phi_e\rangle_0$ by implementing the original DST on the $N_1$ systems. In the second step, based on the results of the first step, we reconstruct a pure state $|\phi_r\rangle$ by operating the revised DST on the $N_2$ systems. Based on the results of the two steps, we could reconstruct a final estimated state $|\phi_f\rangle$ with high efficiency for all unknown states.

In the first step, for the $N_1$ unknown systems, by choosing an appropriate measurement strength $g_1$, we could obtain a estimated state $\rho_e$ by using the original DST~\cite{zhan,zhu1}. From Eq. (\ref{e2}), the $n$th row of the state is constructed by measuring the observable $|n\rangle\langle n|$. Thus, the obtained $\rho_e$ is not hermite. To improve the efficiency, we could optimize the estimated state by
\begin{equation}\label{eop}
\rho_e'=\frac{\rho_e+\rho_e^\dagger}{\mathrm{tr}(\rho_e+\rho_e^\dagger)}.
\end{equation}

However, the state $\rho_e'$ is not a pure state. From the result $\rho_e'$, we could construct a final pure state by
\begin{equation}\begin{split}\label{ecp}
|\phi_e\rangle =\sum_m\left(\sum_n \langle a_m|\rho_e'|a_n\rangle \frac{\langle a_n|\rho_e'|a_n\rangle}{\langle a_1|\rho_e'|a_n\rangle}\right)|a_m\rangle,
\end{split}\end{equation}
where the coefficient $1/\langle a_1|\rho_e'|a_n\rangle$ is used to ensure a uniform common phase. Because the efficiency is very low when $\langle a_n|\phi_s\rangle\langle \phi_s|a_n\rangle \to 0$, as shown in Fig. 1. The weight factor $\langle a_n|\rho_e'|a_n\rangle$ in Eq. (\ref{ecp}) is used to improve the efficiency. The state $|\phi_e\rangle$ is not normalized, we could obtain a pure normalized state by the normalization $|\phi_e\rangle_0=|\phi_e\rangle/\sqrt{\langle \phi_e|\phi_e\rangle}$.

From the results of Fig. 1, the efficiency of the revised DST is high when the measurement operator $A$ satisfies $\langle \phi_s|A|\phi_s\rangle \to 1/d$. Thus, we should search an operator $A$ that satisfies $\langle \phi_s|A|\phi_s\rangle \to 1/d$ to improve the efficiency of the revised DST. Based on obtained the estimated pure state $|\phi_e\rangle_0$, we could construct a orthogonal basis $\{|\phi_e\rangle_0,|\phi_e\rangle_1,...,|\phi_e\rangle_{d-1}\}$ by using Gram-Schmidt procedure. As $|\phi_e\rangle_0$ is the estimated state of the true state $|\phi_s\rangle $, we can construct the state
\begin{equation}\label{cmub}
|a\rangle=\frac{1}{\sqrt{d}}\sum_{m=0}^{d-1}|\phi_e\rangle_i,
\end{equation}
which satisfies $\langle \phi_s|a\rangle\langle a|\phi_s\rangle \to 1/d$.

In the second step, for the other $N_2$ unknown systems, by choosing an appropriate measurement strength $g_2$ and implementing measurement of the observable $A = |a\rangle\langle a|$ given by Eq. ({\ref{cmub}}), we could use the revised DST strategy to reconstruct an estimated state $|\phi_r\rangle$ with high efficiency from Eq. ({\ref{e011}}).

The mean square errors (MSEs) of the first step and the second step could be calculated by Monte Carlo simulations, which are denoted by $\mathcal{E}_1$ and $\mathcal{E}_2$ respectively. From the results of the two steps, we could reconstruct the final estimated pure state by
\begin{equation}\label{ef}
|\phi_f'\rangle=\frac{|\phi_e\rangle_0}{\mathcal{E}_1}+\frac{|\phi_r\rangle}{\mathcal{E}_2},
\end{equation}
where $|\phi_f'\rangle$ is not a normalized state. Thus, the state $|\phi_f\rangle=|\phi_f'\rangle/\sqrt{\langle \phi_f'|\phi_f'\rangle}$ is our final estimated state in the hybrid DST.

\begin{figure}[t]
\centering \includegraphics[scale=0.5]{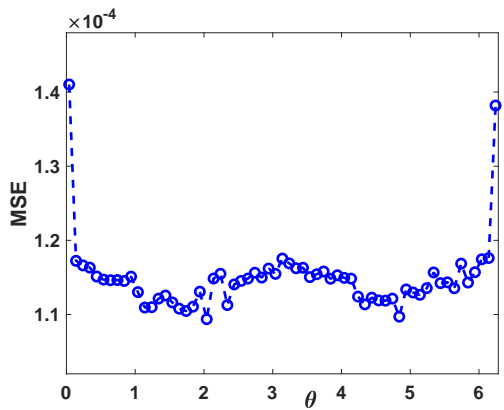} \caption{(Color online) The values of the MSEs are pictured for different unknown states in hybrid DST. All the MSEs have been averaged over $10^4$ repeated simulations. The total number of the quantum systems is $N=2\times 10^4$, the number of the systems used in the original DST is $N_1=4\times 10^3$, and the number of the systems used in the revised DST is $N_2=1.6\times 10^4$. The coupling strengths of the original and the revised DST are $g_1=1.2$ and $g_2=0.4$ respectively. }
\label{fig02}
\end{figure}

To show the useful and high efficiency for all unknown pure states, we perform the Monte Carlo simulations of the hybrid DST. In the simulations, the unknown state is denoted by $|\phi_s\rangle= \cos(\theta/2)|0\rangle+\sin(\theta/2)|1\rangle$. The total number of the unknown quantum systems is $N=2\times 10^4$, the number of the systems for the first step is $N_1=4\times 10^3$, and the number $N_2=1.6\times 10^4$. The coupling strength of the original DST is $g_1=1.2$, and the coupling strength of the revised DST is $g_2=0.4$. The MSEs of the different input states have been pictured in Fig. 2. We can see that the hybrid DST  is useful with high efficiency for all unknown states. In the next section, we will compare the efficiencies of the hybrid DST and the conventional QST.

\section{Comparison}\label{sec5}
In this section, we calculate the MSE of the hybrid DST and present the comparison between the hybrid DST and the conventional QST. One of the best known conventional QST strategies is SU(2) tomography which is based on the best known optical homodyne tomography~\cite{qst001,qst002,qst1,qst2,qst5}. SU(2) tomography is valid for arbitrary dimensional systems. However, in SU(2) tomography, the measurement operators must be chosen as all the bases in Haar¡¯s invariant measure, which is difficulty to be realized.

Besides the SU(2) tomography, there are two well-established state estimation strategies. One is composed of the measurements on a complete sets ($d+1$ sets for $d$-dimensional systems) of mutually unbiased bases (MUBs)~\cite{mubt1,mubt2}, which is denoted as MUB tomography here. The other is composed of symmetric informationally complete (SIC) measurements~\cite{qst11,qst12,qst13}, we denote it as SIC tomography.

The scaled  MSE is a good measure of the efficiency, which is defined by
\begin{equation}\label{smse}
\mathcal{E}_{s}(\rho)=N\mathcal{E},
\end{equation}
where $N$ is the number of the unknown quantum systems.

For pure $d$-dimensional unknown states, the scaled MSE achievable of the MUB tomography is~\cite{qst11,qst13, mubt2}
\begin{equation}\label{emub}
\mathcal{E}_{s}(\rho)=d^2-1.
\end{equation}

For pure unknown states, the scaled MSE for optimal SIC tomography is~\cite{qst11,qst12,qst13}
\begin{equation}\label{sicb}
\mathcal{E}_{s}(\rho)= d^2+d-2.
\end{equation}

We use the Monte Carlo simulations to calculate the scaled MSEs of the SU(2) tomography and the hybrid DST. In hybrid DST, we use $10^4d$ identical unknown systems to calculate the scaled MSE for the $d$-dimensional systems. First, we perform the original DST on the $2\times 10^3d$ systems to obtain a estimated pure state $|\phi_e\rangle_0$ given by Eq. ({\ref{ecp}}). Second, based on the obtained $|\phi_e\rangle_0$, using the revised QST strategy, we reconstruct the estimated state $|\phi_r\rangle$ from the results of the measurements on the $8\times 10^3d$ systems. Finally, the final reconstructed state $|\phi_f\rangle$ could be derived by Eq. (\ref{ef}). In the simulations of the original DST, the coupling strengths of the different dimensions are all equal $g_1=1.2$. And the coupling strengths $g_2$ of the revised DST are listed in the table 1 for different dimensions.
\begin{table}[!htbp]
\renewcommand{\arraystretch}{1.5}

\begin{tabular}{c c c c c c}
\hline
\hline
\multirow{1}{*}{} $d=2,3$ & $d=4,5,6,7,8$ & $d=9$ & $d=10,11,12$ & $d=13,14,15$ \\

\multirow{1}{*}{} $g_2=0.4$ & $g_2=0.6$ & $g_2=0.7$ & $g_2=0.8$ & $g_2=0.9$ \\

\hline
\hline
\end{tabular}

\caption{The coupling strengths $g_2$ for different dimensions in the revised DST.}

\end{table}

\begin{figure}[!htbp]
\centering \includegraphics[scale=0.5]{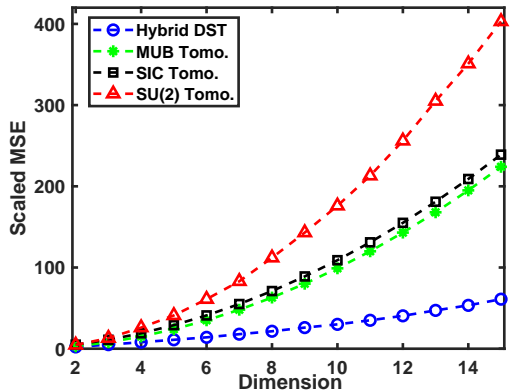} \caption{(Color online) The values of the scaled MSEs are pictured for different tomography strategies. All the MSEs of the hybrid DST and SU(2) tomography have been averaged over $10^3$ different randomly selected unknown states. Circles represent the hybrid DST, stars represent the MUB tomography, squares represent the SIC tomography, and triangles represent the SU(2) tomography.}
\label{fig03}
\end{figure}

All the values of the scaled MSEs for different tomography strategies are pictured in Fig. 3. We can see that the efficiency of the hybrid DST is significantly higher than that of the conventional QST for pure states, especially for the high-dimensional quantum systems. For example, for the 15-dimensional systems, the obtained scaled MSE of the hybrid DST is about $61$, while the scaled MSE of the SIC tomography is $238$. On the other hand side, the hybrid DST is more easily realized in actually experiments. Thus, the hybrid DST may be much more useful in the realization of state tomography.

\section{conclusion}\label{sec6}
We have revised the direct state tomography for pure state. In the revised DST, $N$ times measurements can be accomplished by using only $2N$ identical unknown systems, which is independent on the dimension of the systems. Based on the original DST and the revised DST, we construct the hybrid DST. In the hybrid DST, the efficiency is improved and significantly higher than that of the conventional QST. DST is much easier to be implemented in actual experiments. Thus the hybrid DST may be useful in reconstructing unknown pure quantum states.

\section*{Acknowledgments}
This work was financially supported by the National Natural Science Foundation of China (Grants No. 11965005, No. 11801430, and No. 11664005), X. Zhu was also supported by the Plan Project for Guizhou Provincial Science and Technology (No. QKH-PTRC [2018] 5803) and the Scientific Research Foundation of Guizhou University of Finance and Economics.


\begin{thebibliography}{00}
\section*{References}
\bibitem{niel} M.~A.~Nielsen, I.~Chuang, in \emph{Quantum Computation and Quantum Information}, Cambridge University Press, Cambridge, 2000.
\bibitem{qst001} D. T. Smithey, M. Beck, M. G. Raymer, and A. Faridani, Phys. Rev. Lett. \textbf{70} (1993) 1244.
\bibitem{qst002} G. M. D¡¯Ariano, U. Leonhardt, and M. Paul, Phys. Rev. A \textbf{52} (1995) R1801.

\bibitem{qst1} G.~M.~D'Ariano, L.~Maccone, and M.~Paini, J. OPT. B: Quantum Semicalss. Opt. \textbf{5} (2003) 77.
\bibitem{qst2} G.~M.~D'Ariano, L.~Maccone, and M.~G.~A.~Paris, J. Phys. A \textbf{34} (2001) 93.
\bibitem{qst3} D.~F.~V.~James, P.~G.~Kwiat, W.~J.~Munro, and A.~G.~White, Phys. Rev. A \textbf{64} (2001) 052312.
\bibitem{qst4} A.~E.~Allahverdyan, R.~Balian, and Th.~M.~Nieuwenhuizen, Phys. Rev. Lett. \textbf{92} (2004) 120402.
\bibitem{qst5} G.~M.~D'Ariano, L.~Maccone, and M.~F.~Sacchi, in \emph{Qautnum information with Continuous Variables of Atoms and Light}, edited by N.Cerf, G.~Leuchs, and E. Polzik, World Scientific Press, London, 2007.
\bibitem{qst6} T.~Durt, C.~Kurtsiefer, A.~Lamas-Linares, and A.~Ling, Phys. Rev. A \textbf{78} (2008) 042338.
\bibitem{qst7} R.~B.~A.~Adamson and A.~M.~Steinberg, Phys. Rev. Lett. \textbf{105} (2010) 030406.
\bibitem{qst8} H.~Wang, W.~Zheng, Y.~Yu, M.~Jiang, X.~Peng, and J.~Du, Phys. Rev. A \textbf{89} (2014) 032103.
\bibitem{qst9} A. I. Lvovsky, and M. G. Raymer, Rev. Mod. Phys. \textbf{81} (2009) 299.
\bibitem{qst10} G. M. D'Ariano and P. Perinotti, Phys. Rev. Lett. \textbf{98} (2007) 020403.
\bibitem{qst11} H. Zhu, Ph.D. thesis, National University of Singapore, 2012, available at http://scholarbank.nus.edu.sg/bitstream/handle/10635 /35247/ZhuHJthesis.pdf
\bibitem{qst12} H. Zhu and B.-G. Englert, Phys. Rev. A \textbf{84} (2011) 022327.
\bibitem{qst13} H. Zhu, Phys. Rev. A \textbf{90} (2014) 012115.
\bibitem{qst14} J. \v{R}eh\'{a}\v{c}ek , Y. S. Teo, and Z. Hradil, Phys. Rev. A \textbf{92} (2015) 012108.
\bibitem{qst15} N. Bent, H. Qassim, A. A. Tahir, D. Sych, G. Leuchs, L. L. S\'{a}nchez-Soto, E. Karimi, and R.W. Boyd, Phys. Rev. X \textbf{5} (2015) 041006.
\bibitem{aav} Y.~Aharonov, D.~Z.~Albert, and L.~Vaidman, Phys. Rev. Lett. \textbf{60} (1988) 1351.
\bibitem{wk1} I.~M.~Duck, P.~M.~Stevenson, and E.~C.~G.~Sudarshan, Phys. Rev. D \textbf{40} (1989) 2112.
\bibitem{wk2} R.~Jozsa,Phys. Rev. A \textbf{76} (2007) 044103.
\bibitem{wk3} S.~Wu and Y.~Li, Phys. Rev. A \textbf{83} (2011) 052106.
\bibitem{wk4} X.~Zhu, Y.~Zhang, S.~Pang, C.~Qiao, Q.~Liu, and S.~Wu, Phys. Rev. A \textbf{84} (2011) 052111.
\bibitem{dsm1} J.~S.~Lundeen, B.~Sutherland, A.~Patel, C.~Stewart, and C.~Bamber, Nature (London) \textbf{474} (2011) 188.
\bibitem{dsm2} H.~F.~Hofmann, Phys.Rev. A \textbf{81} (2010) 012103.
\bibitem{dsm3} S.~Wu, Sci. Rep. \textbf{3} (2013) 1193.
\bibitem{dsm4} J.~Z. Salvail, M.~Agnew, A.~S.~Johnson, E.~Bolduc, J.~Leach, and R.~W.~Boyd, Nat. Photon. \textbf{7} (2013) 316.
\bibitem{dsm5} Y.~Shikano, in \emph{Measurements in Quantum Mechanics}, edited by M.~R.~Pahlavani, InTech, Rijeka, 2012, p. 75.
\bibitem{dsm6} Z. Shi, M. Mirhosseini, J. Margiewicz, M. Malik, F. Rivera, and Robert W. Boyd, e-print arXiv: 1503.04713 [quant-ph]
\bibitem{dsm7} M. Mirhosseini, O. S. Maga\~{n}a-Loaiza,S.~M.~HashemiRafsanjani, and R. W. Boyd, Phys. Rev Lette. \textbf{113} (2014) 090402.
\bibitem{mac} L.~Maccone and C.~C.~Rusconi, Phys. Rev. A \textbf{89} (2014) 022122.
\bibitem{zhan} Y-X.~Zhang, S.~Wu, and Z-B.~Chen, Phys. Rev. A \textbf{93} (2016) 032128.
\bibitem{vall} G.~Vallone and D.~Dequal, Phys. Rev. Lett. \textbf{116} (2016) 040502.
\bibitem{gro} J. A. Gross, N. Dangniam, C. Ferrie, and C. M. Caves, Phys. Rev. A \textbf{92} (2015) 062133.
\bibitem{zou}  P.~Zou, Z-M.~Zhang, and W.~Song, Phys. Rev. A \textbf{91} (2015) 052109.
\bibitem{zhu1}  X.~Zhu, Y.~X.~Zhang, and S.~Wu, Phys. Rev. A \textbf{93} (2016) 062304.
\bibitem{zhu2}  X.~Zhu and Q. Wei, Ann. Physics \textbf{376}(2017) 283~295.
\bibitem{mub}  T.~Durt, B.~Englert, I.~Bengtsson, and K. Zyczkowski, Int. J. Quantum Inf. \textbf{8} (2010) 535.
\bibitem{mubt1} W. K. Wootters and B. D. Fields, Ann. Phys. (NY) \textbf{191} (1989) 363.
\bibitem{mubt2} F. Embacher and H. Narnhofer, Ann. Phys. (NY)  \textbf{311} (2004) 220.

\end{thebibliography}
\end{document}